\documentclass[prd, amsfonts, twocolumn, nofootinbib, showpacs]{revtex4-2}
\usepackage{graphicx, epsfig}
\usepackage{color}
\usepackage{amsmath}
\usepackage{amssymb}
\usepackage{physics}
\newcommand{\be}{\begin{equation}}
\newcommand{\ee}{\end{equation}}
\newcommand{\bea}{\begin{eqnarray}}
\newcommand{\eea}{\end{eqnarray}}

\newcommand{\gapp}{\mathrel{\raise.3ex\hbox{$>$}\mkern-14mu
\lower0.6ex\hbox{$\sim$}}}
\newcommand{\lapp}{\mathrel{\raise.3ex\hbox{$<$}\mkern-14mu
\lower0.6ex\hbox{$\sim$}}}
\def\bbox{{\,\lower0.9pt\vbox{\hrule \hbox{\vrule height 0.2 cm
\hskip 0.2 cm \vrule  height 0.2 cm}\hrule}\,}}

\begin{document}
\title{Entropy created when colliding particles fall into a black hole}
\author{De-Chang Dai$^{1,2}$ }
\affiliation{ $^1$ Department of Physics, national Dong Hwa University, Hualien, Taiwan, Republic of China}
\affiliation{ $^2$ CERCA, Department of Physics, Case Western Reserve University, Cleveland OH 44106-7079}

\begin{abstract}
\widetext

If two particles collide in the vicinity of a black hole horizon, their center of mass energy is practically unlimited, so another black hole with a large mass and thus entropy can be created. The resulting black hole can then merge with the original one. If the black hole is created very close to the horizon, its energy will be highly redshifted for an asymptotic observer. However, its entropy is not redshifted. We demonstrated that the newly created entropy can be higher than the Bekenstein-Hawking entropy of the final black hole, though we neglect that a certain amount of energy can escape to infinity, carrying away part of the entropy produced in the process. This is a counter-example to the statement that the black hole thermal entropy counts all the states inside the black hole. Unlike similar examples, this colliding process does not involve exotic matter, alternative theories of gravity, nor artificial ad hoc gluing of two different spacetimes.

\end{abstract}


\maketitle
\section{Introduction}
Black holes are the most interesting solutions of Einstein's equations \cite{Schwarzschild:1916uq}. A black hole solution is characterized by associated thermodynamic quantities like temperature, entropy, energy etc. These quantities are closely connected to quantum effects. For example,  entropy of a black hole is related to its area, i.e. $S_{bh}=A/4$\cite{Hawking:1975vcx,Bekenstein:1973ur}.  A very important question is then whether this entropy counts all the states inside a black hole, which is the topic of black hole holography \cite{tHooft:1993dmi,Susskind:1994vu} and AdS/CFT correspondence \cite{Maldacena:1997re}.     
There are several semiclassical configurations that can carry more entropy than the black hole thermal radiation, i.e. Wheeler's bag of gold\cite{Wheeler:1964}, Monster\cite{Sorkin:1981wd,Hsu:2007dr,Hsu:2008yi}, and Kruskal-FRW gluing  \cite{Wheeler:1964,Hsu:2008yi}.  This clearly implies that the Bekenstein-Hawking entropy is not enough to account for all the states inside a black hole. This fact is closely related to the black hole information loss paradox\cite{Mathur:2009hf}.  
Although these classical configurations give a hint that the black hole surface may not possess all the information about the interior, they do require artificial gluing two different spacetimes or exotic matter. It is therefore well motivated to look for a new example that involves only standard matter and does not invoke manipulations in the form of cutting and pasting. 

\begin{figure}[h]
\includegraphics[width=3cm]{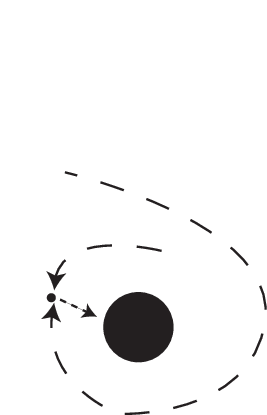}
\caption{Two particles collide near a black hole horizon. The gravitational potential is so strong that the center of mass energy can achieve any possible value. A lighter black hole is formed locally and falls into the original black hole. }
\label{path}
\end{figure}

It has been shown that an extremal rotating black hole can act as a particle accelerator and create extremely high center of mass energy collisions (Fig. \ref{path})\cite{Banados:2009pr}. The same process cannot be applied to a Schwarzschild black hole since two free-falling particles starting from infinity have a small relative velocity when they collide. This idea, however, can be rescued if a low energy particle is escaping from the center\cite{Hackmann:2020ogy}. This process gives a possible mechanism to create a black hole with high entropy through particle collision near the horizon. 

In the present paper, we first discuss the evolution of an apparent horizon when two shock waves collide. Then, we assume that the black hole entropy must be larger than one-fourth of the area of the apparent horizon at any moment. This gives a lower bound of the entropy creation during the collision. A considerable fraction of entropy is created before the object created in this collision hits the singularity and moreover, before its center of mass passes the horizon.  We note that this study neglects that a certain amount of energy can escape to infinity. This energy can also carry away part of the entropy produced in the process. However, we also expect that the radiation should be proportional to the of square the falling object's energy. In some situations, this radiation can be tuned to almost zero. Thus, we demonstrate that the final black hole's horizon may not account for all the entropy at its formation.

\section{Shock wave solution}

Collisions of two relativistic particles have been studied intensively either with analytical \cite{DEath:1978,DEath:1992,DEath:1992-1,DEath:1992-2,DEath:1993yud,Eardley:2002re,Kohlprath:2002yh,Yoshino:2005hi,Kaloper:2007pb,Alvarez-Gaume:2008qeo,Pretorius:2018lfb} or numerical methods\cite{Anninos:1994gp,Anninos:1995vf,1997ASPC..123..314H,Sperhake:2005uf,Loffler:2006wa,Shibata:2008rq,Choptuik:2009ww,East:2011aa,East:2012mb}. It was shown that the energy loss in head-on collisions is less than $1-1/\sqrt{2}$ times the total energy in the analytical studies, and even smaller (0.002 times the total energy) in numerical studies\cite{Anninos:1994gp}. Here we follow the analytic method to study the apparent horizon evolution after two shock waves colliding\cite{Yoshino:2005hi}.
.    

 \begin{figure}[h]
\includegraphics[width=5cm]{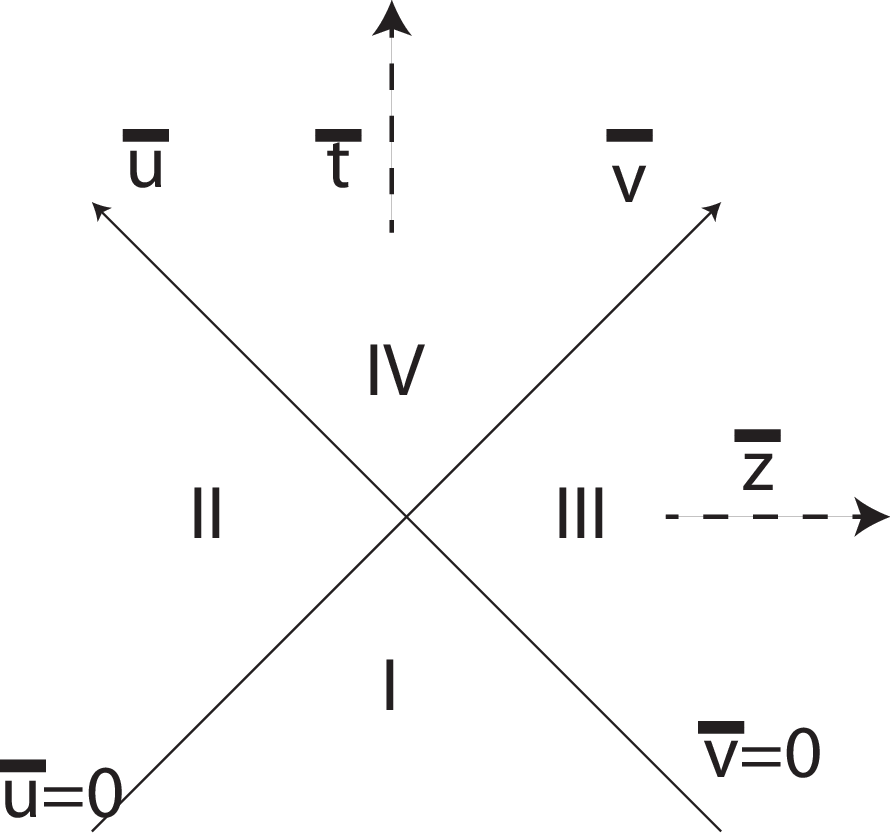}
\caption{Two relativistic particles are moving along the paths $\bar{u}=0$ and $\bar{v}=0$ respectively. They collide at $\bar{u}=\bar{v}=0$ and separate the space-time into regions I, II, III and IV. It is impossible to find an analytical solution in region IV. However, an analytical solution does exist in regions I, II, and III. }
\label{range}
\end{figure}

The paths of two relativistic particles in the Minkowski space $(\bar{t},\bar{x},\bar{y},\bar{z})$ are expressed in terms of two parameters 

\begin{eqnarray}
\bar{u}&=&\bar{t}-\bar{z}\\
\bar{v}&=&\bar{t}+\bar{z}
\end{eqnarray}
The paths of these two particles are $\bar{u}=0$ and $\bar{v}=0$ respectively (Fig. \ref{range}). These two particles collide at $\bar{u}=\bar{v}=0$. It is impossible to find an analytic solution in region IV. However, we are looking for the formation of an apparent horizon, so a complete solution is not needed. Regions II and III are particularly interesting since they represent the  space-time after the collision. Regions II and III are mirror symmetric if these two particles are identical except for their momenta being opposite.  Therefore, we are going to construct the solution in Region II. The quantities in Region III can be found easily from there. Here, energy of the particle is $\mu=\gamma m_0$, where $m_0$ and $\gamma$ are particle mass and Lorentz factor.  

 The metric in region II is considered to be a shock wave solution, as proposed by Aichelburg and Sexl\cite{1971GReGr...2..303A},  

\begin{equation}
ds^2=-d\bar{u}d\bar{v}-2\ln \bar{\rho} \delta (\bar{u}) d\bar{u}^2 +d\bar{\rho}^2+\bar{\rho}^2 d\bar{\phi}^2
\end{equation}
Here, the units of length are chosen to be $r_0=4 G \mu$. The $\delta$-function in the metric implies that a shock wave is at $\bar{u}=0$.  To avoid the discontinuity at $\bar{u}=0$, a smooth and continuous coordinate is chosen via the coordinate transformation,
\begin{eqnarray}
\bar{u}&=&u \\
\bar{v}&=&v-2\theta(u)\ln(\rho)+\frac{u\theta (u)}{\rho^2} \\
\bar{\rho}&=& \rho\Big(1-\frac{u\theta(u)}{\rho^2}\Big)\\
\bar{\phi}&=&\phi
\end{eqnarray}
where $\bar{\rho}=\sqrt{\bar{x}^2+\bar{z}^2}$. The metric becomes 
\begin{eqnarray}
ds^2&=&-dudv+\Big(1+\frac{u\theta(u)}{\rho^2}\Big)^2d\rho^2\nonumber \\
&+&\Big(1-\frac{u\theta(u)}{\rho^2}\Big)^2\rho^2 d\phi^2 .
\end{eqnarray}

\begin{figure}[h]
\includegraphics[width=5cm]{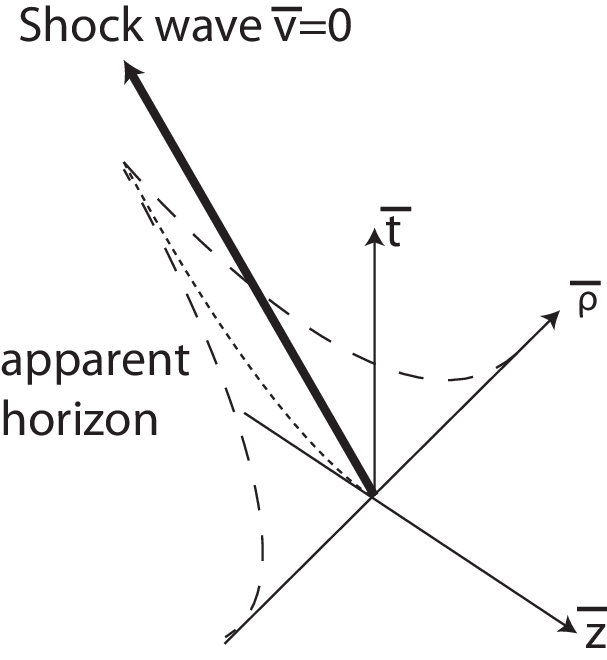}
\caption{The apparent horizon is formed after two particles  collide. The long-dashed lines represent the apparent surface in Minkowski space. The short-dashed line represents the center of the surface at a particular moment $\bar{t}$. The surface of the apparent horizon is ahead of the shock wave (the thick arrow).}
\label{collision}
\end{figure}

Since we are interested in the apparent horizon location, we follow \cite{Yoshino:2005hi} and define a function $h(\rho)$ for the apparent horizon after a head-on collision as
\begin{eqnarray}
\label{u1}
u&=&h(\rho)\\
\label{v1}
v&=&0 ,
\end{eqnarray}
where $\bar{u}$ is mapped to $\rho^2 \ge h(\rho)\ge 0$. $h(\rho)$ is found by solving the apparent horizon equation\cite{Yoshino:2005hi}. The solution for the head-on collision  is 
\begin{eqnarray}
&&h=2\rho^2 \ln (\rho)\\
&&\rho_{min}=1\\
&&\rho_{max}=\sqrt{e} .
\end{eqnarray} 
 The effective metric on the apparent horizon at $v=0$ is written as  
\begin{equation}
ds_{eff}^2 = \Big(1+2\ln(\rho)\Big)^2d\rho^2+\Big(1-2\ln(\rho)\Big)^2\rho^2 d\phi^2 .
\end{equation}
By integrating, one finds the apparent horizon area,  
\begin{equation}
A_{v=0}=\int \Big( 1-4\ln^2(\rho)\Big)\rho d\rho d\phi =\pi r_0^2
\end{equation}
$r_0$ is reinstalled. There is a mirror symmetric apparent horizon at Region III. The total apparent horizon area is 
\begin{equation}
A=2\pi r_0^2
\end{equation}
These two apparent horizons are obtained in the ${u,v}$ coordinates. We have to transform them back to $(\bar{t},\bar{x},\bar{y},\bar{z})$ coordinates.
In $(\bar{t},\bar{x},\bar{y},\bar{z})$ coordinates the corresponding time and location of the horizon are
\begin{eqnarray}
\bar{t}&=&(\rho^2-1) \ln(\rho)\\
\bar{z}&=&-(\rho^2+1) \ln(\rho) .
\end{eqnarray}
One finds $\frac{e-1}{2}\ge \bar{t}\ge 0$ and $0\ge \bar{z}\ge - \frac{e+1}{2}$. The maximum of $\bar{t}$ is $\bar{t}_{max}=\frac{e-1}{2}$ and the maximum of $\bar{z}$ is $\bar{z}_{min}=-\frac{e+1}{2}$. Thus, the apparent horizon is ahead of the shock wave at any moment $\bar{t}$. Fig.\ref{collision} shows the apparent horizon in $(\bar{t},\bar{x},\bar{y},\bar{x})$ coordinates.

\begin{figure}[h]
\includegraphics[width=8cm]{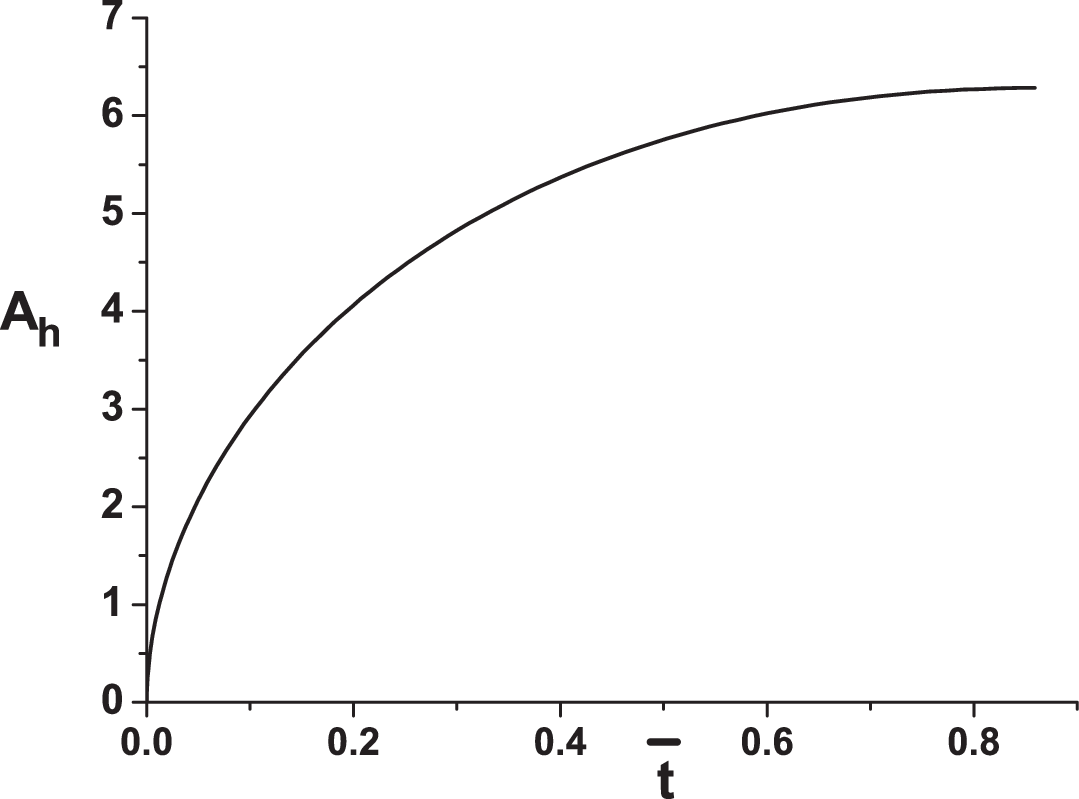}
\caption{The lower bound on the area of the apparent horizon is a monotonically increasing function. It achieves maximum at $\bar{t}_{max}$.}
\label{area-time}
\end{figure}

This analytic solution is incomplete since it does not have a closed apparent surface at any moment $\bar{t}$. Therefore, we assume that once the horizon is formed, its area cannot shrink subsequently. Based on this assumption, a limitation of the horizon area is   
\begin{eqnarray} 
A_h &\ge& 2 \int_0 ^{2\pi}\int_0^{\rho(\bar{t})} (1-2\ln(\rho))(1+2\ln(\rho))\rho d\rho d\phi  \nonumber\\
&=& 2\pi\Big(1-\rho(\bar{t})^2(1+4 \ln^2 (\rho(\bar{t}))-4  \ln(\rho(\bar{t})) \Big) 
\end{eqnarray}
where the factor $2$ is introduced because both of regions II and III must be included. The lower bound of the horizon area is shown in Fig.\ref{area-time}. It is a monotonically increasing function.

\section{Free fall}

After the collision, the resulting configuration must be treated as a single object. This object then falls, crosses the horizon, and hits the singularity. To study the possibility of having an event that generates more entropy than the final horizon can accommodate, the mass of the final black hole must be minimized. Therefore, we assume the momentum of the object is zero right after collision. The object then falls along the radial direction. 

The metric for the Schwarzschild black hole is

\begin{equation}
d\tau^2=\Big(1-\frac{2M}{r}\Big)dt^2-\frac{1}{1-\frac{2M}{r}} dr^2-r^2d\Omega^2 .
\end{equation}
Here, $M$ and $\tau$ are the mass of the black hole and proper time respectively. The equation of motion of a massive particle is   
\begin{equation}
\Big(1-\frac{2M}{r}\Big) \Big(\frac{dt}{d\tau}\Big)^2-\frac{1}{1-\frac{2M}{r}} \Big(\frac{dr}{d\tau}\Big)^2=1 .
\end{equation}
The energy, $E$, of the particle is conserved while the particle is infalling,  so we have
\begin{equation}
\Big(1-\frac{2M}{r}\Big) \frac{dt}{d\tau}=\epsilon ,
\end{equation}
where $m$ is the particle mass and $\epsilon$ is energy to mass ratio, i.e. $\epsilon=E/m>0$. The proper falling time is 
\begin{eqnarray}
\tau &=& \frac{2M}{1-\epsilon^2} ( \frac{\pi}{2}-\alpha +\frac{1}{2} \sin^2 2\alpha)\\
\alpha &=& \sin^{-1}\sqrt{\frac{r}{r_i}} 
\end{eqnarray}
where $r_i=\frac{2M}{1-\epsilon^2}$ is the collision location. The falling time is set to be $\tau=0$ as $r=r_i$. 
An object arrives at the singularity, $r=0$, at $\tau=\tau_f=\frac{r_i \pi}{2}$. 

When two relativistic particles collide at $r=r_i$, it takes about $\bar{t}_{max}$ to form a horizon. The object must not hit the singularity before the horizon forms.  Hence, $\tau_f>\bar{t}_{max}$, or
\begin{equation}
r_i >\frac{4 (e-1)}{\pi}\mu. 
\end{equation}

$r_i\ge 2M$. This equation reduces to 
\begin{equation}
M>\frac{2 (e-1)}{\pi}\mu
\end{equation}
This equation is satisfied automatically since $M\gg \mu$ in this study. Therefore, these two particles have enough time to create a horizon. 

Two-particle-collision events must be studied in a local Minkowski frame. We write down the local Minkowski frame in Painleve-Gullstrand coordinates\cite{Painleve,Gullstrand:1922tfa,Kanai:2010ae,Toporensky:2022nic} as
\begin{equation}
ds^2=-d\tau^2 +\frac{1}{\epsilon^2} (dr+vd\tau)^2 +r^2d\Omega^2 ,
\end{equation}
where, $v=\sqrt{\epsilon^2-1+\frac{2M}{r}}$ is the local velocity of the object.

\begin{figure}[h]
\includegraphics[width=8cm]{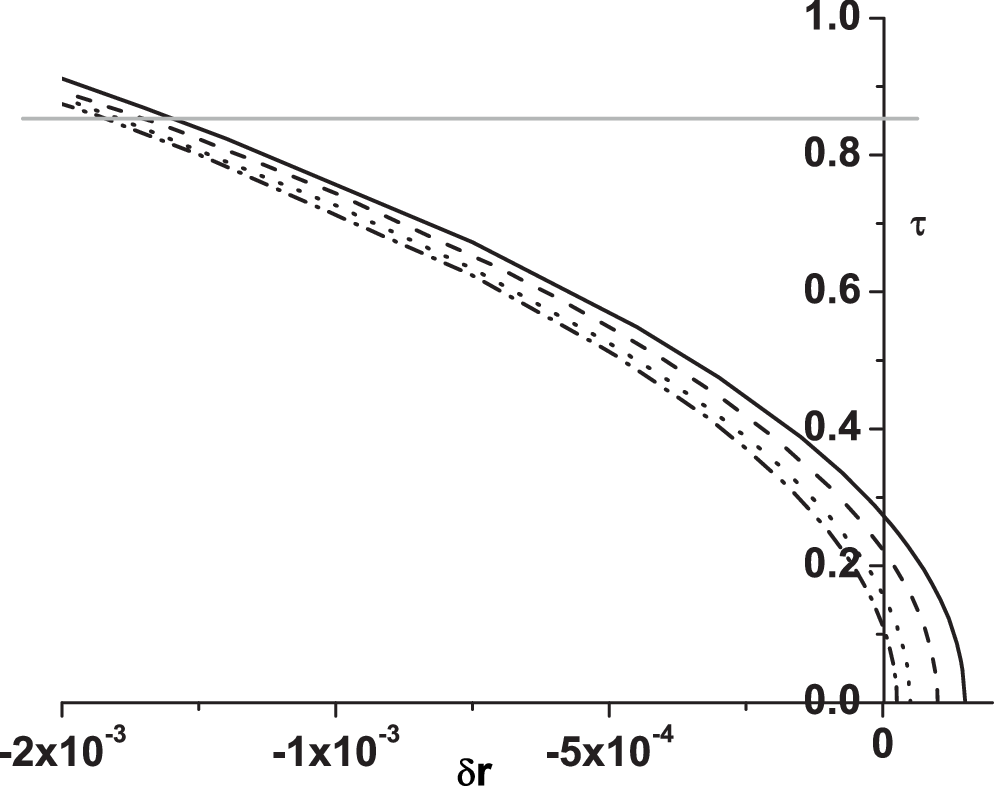}
\caption{$\tau$ vs. $\delta r=r-2M$: $\delta r=0$ represent the black hole horizon. $M=1$, $\mu=10^{-3}$. The solid, dash, dot, and dot-dash curves represent $r_i-2M=6\times 10^{-7}$, $r_i-2M=4\times 10^{-7}$. $r_i-2M=2\times 10^{-7}$, and $r_i-2M= 10^{-7}$ respectively. The gray line represents the $\bar{t}_{max}$. }
\label{time-radius}
\end{figure}

Fig.\ref{time-radius} shows the relation between proper falling time and the radius. The low bounce achieves its maximum at $\bar{t}_{max}$ while the object is still very close to the horizon. Therefore, the colliding object must have settled into a static black hole far before reaching the singularity. The assumption that entropy equals the horizon area divided by $4$ is a reasonable estimate.

\section{Requirement for the collision to generate more entropy than the horizon can possess}

Let's start with two particles where each of them have energy $\mu$ locally, so their total energy is $2\mu$. After including the gravitational potential, the energy of the collision system is   
\begin{equation}
 E=2\mu \sqrt{1-\frac{2M}{r_i}} \ge \Delta M .
\end{equation}
The increased mass of the black hole after collision, denoted as $\Delta M$, is a key factor in our argument. If there is no energy escaping to infinity, then $ \Delta M=E$. 
The area of the final black hole is $4\pi (2M+2\Delta M)^2$. The area of the initial black hole and the apparent horizon is  $4\pi (2M)^2+A_h$. We are looking for the condition for the entropy of the final black hole to be less than the entropy produced in the process. The requirement is    
\begin{equation}
4\pi (2M+2\Delta M)^2 < 4\pi (2M)^2+A_h .
\end{equation}
 We do not know how much energy is escaping to infinity. Hence, we replace this equation with  
\begin{equation}
4\pi (2M+2E)^2 < 4\pi (2M)^2+A_h .
\end{equation}
Since $A_h \ge 2\pi (4\mu)^2$ as $\bar{t}>\bar{t}_{max}$, this relation is held if    
\begin{equation}
\label{r-required}
\frac{\mu}{r_i}>\frac{1}{2}\sqrt{1-\frac{2M}{r_i}}+\frac{\mu }{4M} ,
\end{equation}
 or more precisely 
\begin{equation}
r_i< \frac{M-\frac{\mu^2}{M}+\sqrt{M^2+2\mu^2}}{1-\frac{\mu^2}{4M^2}} .
\end{equation}
This can happen only if $r$ is very close to the horizon. Approximately, it must be within  
\begin{equation}
r_i-2M \lessapprox \frac{\mu^2}{2M} ,
\end{equation}
where $\delta r<<2M$. The entropy created in the process can be greater than the final entropy under this condition.

Based on this requirement, the difference between the final and initial entropies is 
\begin{equation}
\Delta S \leq \pi (2M+2E)^2 - \pi (2M)^2- A_h/4 
\end{equation}

\begin{figure}[h]
\includegraphics[width=8cm]{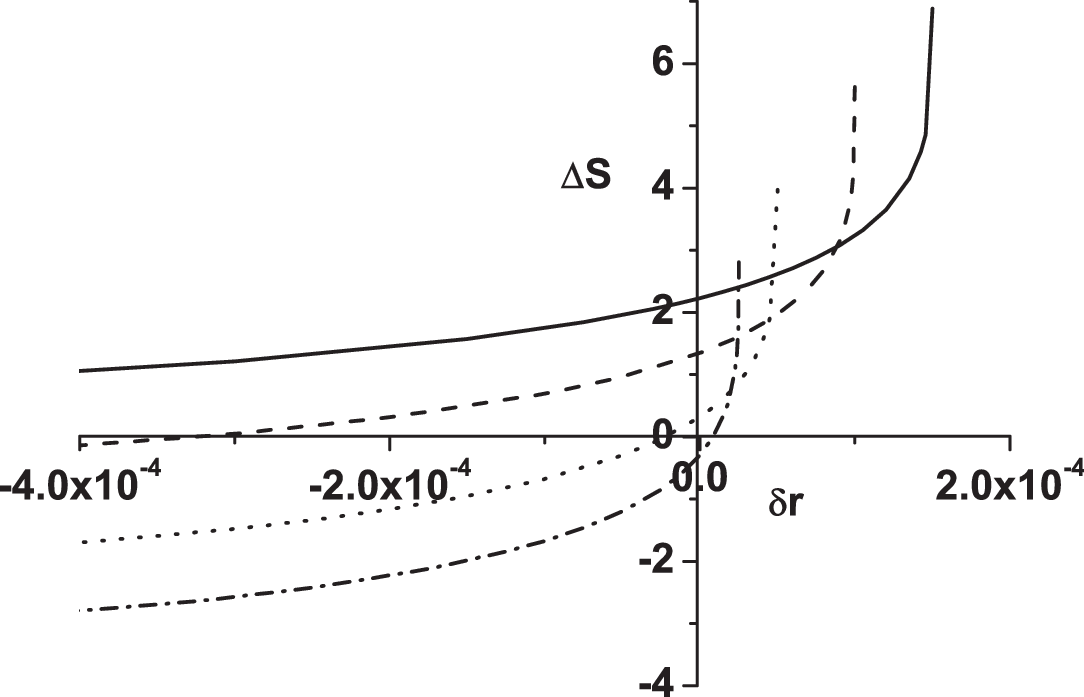}
\caption{The upper bound of $\Delta S$ vs. $\delta r=r-2M$: $\delta r=0$ represent the black hole horizon. $M=1$, $\mu=10^{-3}$. The solid, dash, dot, and dot-dash curves represent $r_i-2M=6\times 10^{-7}$, $r_i-2M=4\times 10^{-7}$. $r_i-2M=2\times 10^{-7}$, and $r_i-2M= 10^{-7}$ respectively.  }
\label{entropy-radius}
\end{figure}

Fig.\ref{entropy-radius} shows the upper bound on $\Delta S$. The solid line shows that if $r_i-2M$ is large enough, the upper bound of $\Delta S$ is larger than $0$, and the final state has large entropy. However, if $r_i-2M$ is small enough that the upper bound of $\Delta S$ is smaller than $0$, and the final state has less entropy than the amount generated during the process. Moreover, the created entropy can be larger than the final state entropy before the center of the mass crosses the horizon if $r_i-2M$ is very small (dot-dashed line in Fig.\ref{entropy-radius}). We note that even when the center of mass has not crossed the horizon yet, the object's apparent horizon is connected to the black hole horizon from an outside observer's point of view. An outside observer cannot see the entire apparent horizon to witness  that such a large amount of entropy is created since the black hole horizon covers this information.

\section{Conclusion}

It is widely accepted  that a black hole entropy is represented by its horizon area. However, it is still not clear whether the Bekenstein-Hawking entropy counts all the states inside the black hole or only states distinguishable from the outside\cite{Marolf:2008tx}. This question is closely connected with the development of the principle of the holography \cite{tHooft:1993dmi,Susskind:1994vu} and AdS/CFT correspondence\cite{Maldacena:1997re}. 
There are a couple examples in which a black hole can have higher inner entropy than the Bekenstei-Hawking entropy, most notably the Wheeler's bag of Gold and Monster\cite{Wheeler:1964,Sorkin:1981wd,Hsu:2007dr,Hsu:2008yi}. A downside of these constructs is that they require either artificial gluing of two different spaces or presence of exotic matter. In the present paper we constructed  a more natural process to embed more entropy inside a black hole than its surface can encode. 

We considered two relativistic particles colliding near the black hole horizon, forming another black hole, and then falling into the horizon. These two relativistic particles can possess an unlimited amount of center of mass energy, so there is no limit on the newly created local entropy. If the black hole is created very close to the horizon, its energy will be highly redshifted for an asymptotic observer. However, its entropy is not redshifted, so the gain in the black hole area (which is the measure of the Bekenstein-Hawking entropy) is not enough to accommodate the entropy gain.   We found that there is enough time for these two particles to form an apparent horizon, which possesses more entropy than the entropy the final black hole after the object hits the singularity (Fig.\ref{entropy-radius}). Thus, a black hole actually could absorb more entropy than its Bekenstein-Hawking entropy. There is no clear answer to where this entropy goes after an object reaches and is destroyed by the singularity. It is possible that the non-linear effects play some important role in this process since this study assumes that a free-falling object is equivalent to an object in an inertial frame. Also, we neglected that a certain amount of energy can escape to infinity. This energy can also carry away a part of the entropy produced in the process. However, we also expect the radiation to be proportional to the square of the falling object's energy. In some situations, this radiation can be tuned to almost zero. Thus, we demonstrated that the final black hole's horizon may not account for all the entropy at its formation.

\begin{acknowledgments}
D.C. Dai is supported by the National Science and Technology Council (under grant no. 111-2112-M-259-016-MY3).    
\end{acknowledgments}

\end{document}